\documentclass[10pt]{article}
\usepackage{amsmath}
\usepackage{amssymb}
\usepackage{graphicx}
\usepackage{cite}
\usepackage[labelfont=bf,labelsep=period,justification=raggedright]{caption}

\makeatletter
\renewcommand{\@biblabel}[1]{\quad#1.}
\makeatother

\date{}

\begin{document}

\begin{flushleft}
{\Large
\textbf{Adaptive and Bounded Investment Returns Promote Cooperation in Spatial Public Goods Games}
}\sffamily
\\[3mm]
\textbf{Xiaojie Chen,$^{1,\ast}$ Yongkui Liu,$^{2,3,4}$ Yonghui Zhou,$^{5}$ Long Wang,$^{6}$ Matja{\v z} Perc$^{7,\S}$}
\\[2mm]
{\bf 1} Evolution and Ecology Program, International Institute for Applied Systems Analysis, Laxenburg, Austria
{\bf 2} School of Automation Science and Electrical Engineering, Beihang University, Beijing, China
{\bf 3} School of Electronic and Control Engineering, Chang'an
University, Xi'an, China
{\bf 4} Center for Road Traffic Intelligent Detection and Equipment
Engineering, Chang'an University, Xi'an, China
{\bf 5} School of Mathematics and Computer Science, Guizhou Normal
University, Guiyang, China
{\bf 6} State Key Laboratory for Turbulence and Complex Systems,
College of Engineering, Peking University, Beijing, China
{\bf 7} Department of Physics, Faculty of Natural Sciences and Mathematics, University of Maribor, Slovenia
\\[2mm]
$^{\ast}$chenx@iiasa.ac.at\\
$^{\S}$matjaz.perc@uni-mb.si
\end{flushleft}
\sffamily

\section*{Abstract}
The public goods game is one of the most famous models for studying
the evolution of cooperation in sizable groups. The multiplication
factor in this game can characterize the investment return from the
public good, which may be variable depending on the interactive
environment in realistic situations. Instead of using the same
universal value, here we consider that the multiplication factor in
each group is updated based on the differences between the local and
global interactive environments in the spatial public goods game,
but meanwhile limited to within a certain range. We find that the
adaptive and bounded investment returns can significantly promote
cooperation. In particular, full cooperation can be achieved for
high feedback strength when appropriate limitation is set for the
investment return. Also, we show that the fraction of cooperators in
the whole population can become larger if the lower and upper limits
of the multiplication factor are increased. Furthermore, in
comparison to the traditionally spatial public goods game where the
multiplication factor in each group is identical and fixed, we find
that cooperation can be better promoted if the multiplication factor
is constrained to adjust between one and the group size in our
model. Our results highlight the importance of the locally adaptive
and bounded investment returns for the emergence and dominance of
cooperative behavior in structured populations.

\section*{Introduction}

The emergence of cooperation among selfish individuals is an
intensively studied problem \cite{Science06Nowak, PR07Szabo}.
Traditionally, the problem of cooperation is investigated by means
of the game theoretical models of the prisoner's dilemma for
pairwise interactions, and more generally public goods game for
groups of interacting individuals. In particular, the public goods
game is abundant in human society, e.g., protecting the global
climate and avoiding overfishing of the oceans \cite{PNAS08Milinski,
PNAS11Santos, EPL11Greenwood, PNAS11Heitzig}. In the classical
public goods game (PGG), individuals engage in multiplayer
interactions and decide simultaneously whether to contribute
(cooperate) or not (defect) to a common pool. Then the accumulated
contributions by cooperators are multiplied by a factor large than
one, i.e., the so-called multiplication factor, and finally the
resulting assets are shared equally among all group members
irrespective of their initial decision. From the
perspective of each individual, defection is clearly the rational
decision to make as it yields the highest income compared to
other members. Thus, selfish individuals should decline to
contribute and attempt to free ride on the other players'
contributions. However, if nobody decides to invest, the group fails
to harvest the benefits of a collective investment, which drives the
population into the tragedy of the commons \cite{Science68Hardin}.
Actually, the group is most successful if everybody
cooperates, and hence the dilemma is caused by the selfishness of individual players.

To study the social dilemma in realistic situations, in the last
decade the risk PGG \cite{PNAS08Milinski, PNAS11Santos}, the
optional PGG \cite{JTB02Hauert, PRL02Szabo, Science02Hauert,
PRSB06Hauert, PRSB07Sasaki}, the threshold PGG \cite{PRE09Wang,
PRE10Szolnoki, BMCEB10Boza, Evolution11Archetti, PLOSONE11Deng}, the
continuous PGG \cite{JTB06Janssen, JTB08Kamimura, JTB11Cressman},
and the ecological PGG \cite{JTB07Wakano, PNAS09Wakano} have been
developed based on the classical PGG from the viewpoint of realistic
societies. On the other hand, several mechanisms for the evolution
of cooperation in the PGG, such as punishment \cite{PRSB03Brandt,
PNAS05Fowler, PNAS06Brandt, Science07Hauert, PNAS09Traulsen,
Nature10Sigmund, PLOSCB10Helbing, PRE10Short, PRE11Szolnokia,
PRE11Szolnokib, PNAS11Baldassarri}, reward \cite{PNAS01Sigmund,
Science09Rand, EPL10Szolnoki, JTB10Hauert, JTB11Sasaki}, reputation
\cite{PNAS01Sigmund, JTB10Hauert}, network reciprocity
\cite{PRE07Guan, EPL08Huang, Nature08Santos, EPL09Wu, EPL09Rong,
PRE09Szolnokia, PRE09Wu, PRE09Yang, PNAS10Fowler, EPL10Shi, EPL10Xu,
PRE10Rong, PRE11Szolnokic, JTB11Vukov, PRE11Perc,
Chaos11Gomez-Gardenes, EPL11Gomez-Gardenes} have been justified. In
particular, complex interaction networks provide a natural and
reasonable framework for studying the PGG in structured populations.
Within this framework, some aforementioned mechanisms, such as
punishment and reward have been further studied \cite{EPL10Szolnoki,
PLOSCB10Helbing, PRE11Szolnokia, PRE11Szolnokib}. Also, some other
factors have been incorporated, such as noise \cite{PRE09Szolnokia},
social diversity \cite{PRE07Guan, Nature08Santos, PRE09Yang,
EPL10Shi}, and success-driven distribution \cite{PRE11Perc}. It is
found that social diversity associated with the number and the size
of the public goods game as well as the individual contribution to
each game can greatly promote the emergence of cooperation
\cite{Nature08Santos}. Indeed, social diversity by means of the
system's other feature information, e.g., game payoffs
\cite{PRE08Perc, EPL10Shi}, teaching activity \cite{PRE07Guan} and
preferential selection \cite{PRE09Yang} in strategy updating, have
been also demonstrated to facilitate cooperation in the PGG.

It is worth mentioning that the inhomogeneities and social diversity
about features of the system are widely existent in human society
and animal world, which can characterize the asymmetric and
different influence of individuals or interacting environments.
However, they are introduced artificially in some previous studies
mentioned above. Indeed the inhomogeneities or social diversity can
emerge spontaneously via the coevolutionary rules, since the values
of property should be not invariable, but evolve based on the state
of the system. In the context of evolutionary game theory, the
adaptive features are often coupled with the strategy evolution. The
coevolution of strategies and features of the model, e.g.,
individual social ties (for example see Refs. \cite{PRE02Ebel,
PRE04Zimmermann, PRE05Zimmermann, PRL06Pacheco, PLOSCB06Santos,
PLOSONE08Poncela, NJP09Poncela, LNCS08pest, LNCS10pest}), noise
level \cite{EPL09Szabo, PRE09Szolnokib}, payoff matrices
\cite{PRE02Tomochi, EPL08Fort}, capability of strategy transfer
\cite{PRE06Wu, NJP08Szolnoki} and individual learning rules \cite{JTB09Moyano, NJP10Cardillo}, have been investigated in different evolutionary games, especially the
prisoner's dilemma game (see \cite{BS10Perc} for a review).
Remarkably, recently Lee et al. further proposed a multiadaptive
prisoner's dilemma game where both the payoff matrices and the
interaction structure are shaped by the behavior of the agents, and
found that such multiadaptive mode can result in the coemergence of
hierarchical structure and cooperation \cite{PRL11Lee}.

At present, we propose a coevolutionary rule in the PGG. We consider
that each interacting group has its own multiplication factor, which
evolves based on the local strategy distribution in the group and
the global strategy distribution in the whole system. Different from
the setting in some previous works \cite{PRE02Tomochi, PRE11Perc,
EPL08Fort, PRL11Lee}, this adaptive multiplication factor is used to
measure the local interacting or cooperative environment in each
group, rather than individual's feature or the whole system's
interaction conditions. Correspondingly, the multiplication factor
represents the feedback return of the local investment to the public
good, and a larger value of the multiplication factor enables a
better investment return \cite{JTB06Janssen}. Structured populations provide a competent framework to describe this local feature, which is updated based on the local and the global level of cooperation. In the present study, in order to focus solely on the effects of the adaptive investment returns, we employ a square lattice where the number of group members is fixed and always the same. Due to
the diversity of local strategy distribution, the investing
cooperators can get together in some groups, whereas in other groups
they are sparse. The inhomogeneous distribution can induce different
investment returns in different interacting groups, which may correspond to the phenomenon of uneven regional exploitation of the common resources in a society. For example, in some pastures the herdsmen may over-exploit the pasture resource by
adding more and more animals to their herd, which may lead to the
gradual desertification of the grassland. Correspondingly, the
socioeconomic returns from herding in these grazing areas decrease
gradually and finally the economic losses are unavoidable. On the
contrary, in other pastures the herdsmen may still use the grassland
resources while at the same time considering the conservation of the
ecosystem, and in such cases a higher socioeconomic return is likely.
We adopt the state of the system, i.e., the global
cooperation level, as the criterion to measure whether the local
cooperative environment is favorable or not. Since the state of the
system is evolving simultaneously, here we prefer the dynamical
global cooperation level, instead of the static criterion in Refs.
\cite{PRE02Tomochi, PRL11Lee}. Moreover, in real situations the
investment return is variable, but should be somewhat limited by
external adjustment. In general, it should be limited in a certain
range \cite{JFE04Kogan}.

In this study, we assume that the enhancement factor in each group
is updated based on the differences between the dynamical local and
global cooperation levels, and limited between the lower $R_l$ and the upper $R_u$ limit. We study how the adaptive and bounded investment returns influence the evolution of
cooperation in the spatial PGG. We find that this PGG model with the
adaptive and bounded investment returns can effectively enhance
cooperation in spatially structured populations, and that appropriately bounded
limitations of the multiplication factor can result in the best
cooperation level. We find further that, in comparison to the
traditionally spatial PGG where only one invariable multiplication
factor which is larger than one exists in the whole population, our
proposed PGG model can produce a higher cooperation level when each
multiplication factor is limited to change only between one and the
group size.

\section*{Results}

\begin{figure}[!ht]
\begin{center}
\includegraphics[width=5in]{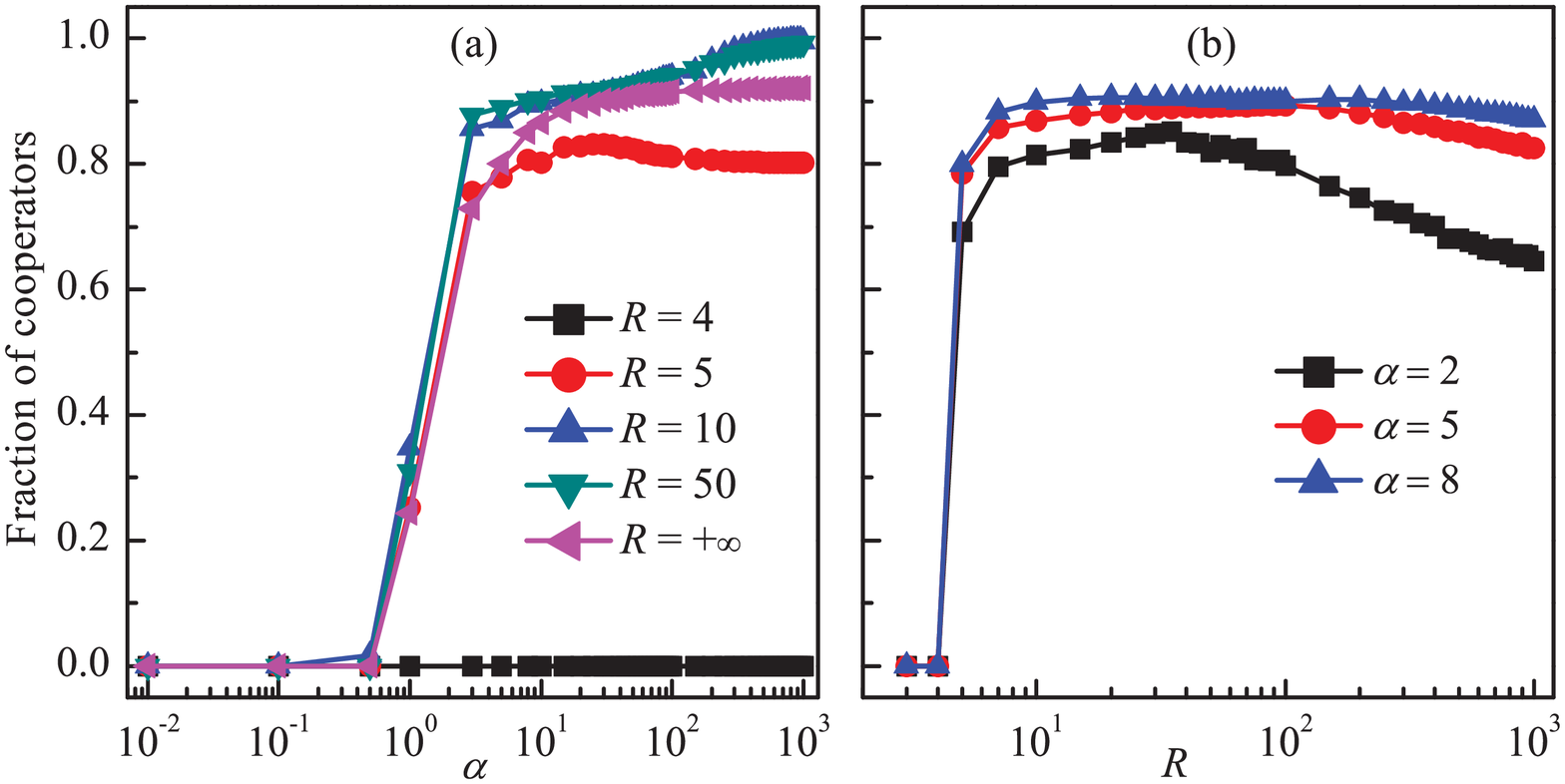}
\end{center}
\caption{ {\bf Promotion of cooperation due to adaptive and bounded
investment returns.} Panel (a) depicts the fraction of cooperators
$\rho$ in dependence on the feedback strength $\alpha$ for different
values of $R$. Panel (b) depicts the fraction of cooperators in
dependence on the boundary value $R$ for different values of
$\alpha$. It can be observed that cooperation can be promoted for
large values of feedback strength, and there exist moderate boundary
values warranting the best promotion of cooperation. Here, $r_0=1$.}
\label{fig1}
\end{figure}

We start by presenting the results as obtained when the
lower limit of the multiplication factor $R_l$ is equal to the opposite
number of the upper limit $R_u$, i.e., $R_u=-R_l=R>1$. Here $R$ is
the limit value (for the detailed definitions see the Methods
section). Figure~\ref{fig1}(a) shows the cooperation level $\rho$ at
equilibrium in the population in dependence on the feedback strength
$\alpha$ for different values of $R$. We find that when $\alpha>1$
cooperation can be promoted for larger $R$. To be specific, when $R$
is small, e.g., $R=4$, full defection is achieved, irrespective of
the values of feedback strength. This is because cooperators
cannot survive in structured populations if the maximum enhancement
factor is not sufficiently large \cite{PRE09Szolnokia}. When $R$ becomes
larger, e.g., $R=5$, the cooperation level first increases and then
slightly decreases. Subsequently, it holds at about $0.8$ with
increasing $\alpha$. While there is no limitation for the
multiplication factor, although the cooperation level is very high,
but a small amount of defectors can survive in the population even
for high feedback strength, e.g., $\alpha=1000$. When moderate
values of $R$ is set, e.g., $R=10$, full cooperation can be achieved
for high feedback strength, and the cooperation level is very
similar for other moderate values of $R$. To further qualify the
effects of $R$ on the evolution of cooperation, we present $\rho$ as
a function of $R$ for different $\alpha$ in Fig.~\ref{fig1}(b).
Clearly, we see that for different values of $\alpha$, the
cooperation level first increases dramatically from zero until
reaching the maximum value at a moderate $R$, then decreases slowly
with increasing $R$. Here, we do not show the cooperation level in
dependence on $R$ for large $\alpha$. In fact, we can still observe
the nonmonotonous dependence of $\rho$ on $R$ even for large
$\alpha$. These results suggest that the PGG model with the adaptive
and bounded multiplication factor can effectively enhance
cooperation in spatial structures, and higher cooperation level can
emerges if an appropriate limitation is considered for the dynamical
multiplication factor.

\begin{figure}[!ht]
\begin{center}
\includegraphics[width=5in]{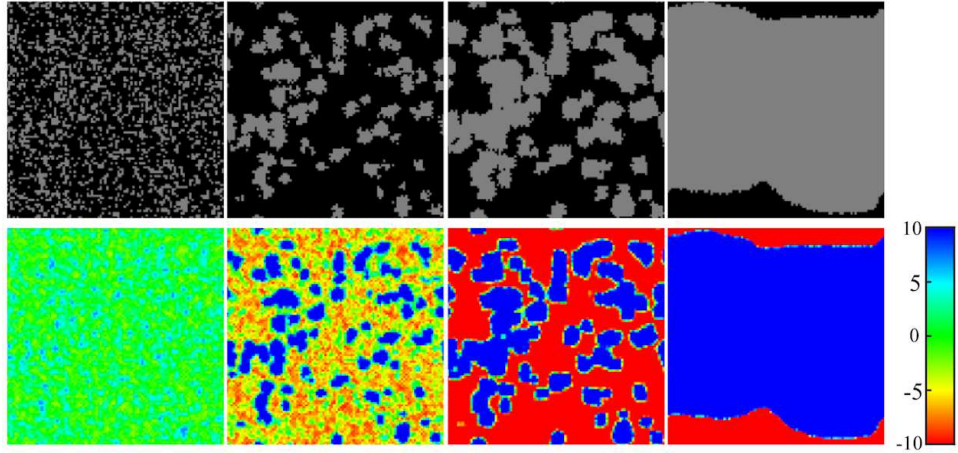}
\end{center}
\caption{ {\bf Characteristic snapshots of strategy and
multiplication factor distributions on a square lattice during the
coevolutionary process.} Top row depicts the time evolution (from
left to right) of typical distributions of cooperators (grey) and
defectors (black) on a square lattice, and bottom row depicts the
corresponding time evolution (from left to right) of typical
distributions of multiplication factor. Results in all panels are
obtained for $\alpha=5$, $r_0=1$, and $R=10$. We have checked that
similar results can emerge for other parameter settings.}
\label{fig2}
\end{figure}

In order to intuitively understand the evolution of cooperation, we
show some typical snapshots of the distribution of strategy and
multiplication factor in the whole population in Fig.~\ref{fig2}. We
find that at the beginning of evolution, cooperators can form many
small and isolated patches. But subsequently, some small compact
cooperator clusters are embedded in the sea of defectors. As time
increases, the cooperator clusters increase gradually, and finally
cooperators may expand as a single ever growing cluster [upper row
in Fig.~\ref{fig2}]. Correspondingly, the multiplication factor in
the full cooperation group can reach the upper limit, whereas the
multiplication factor in the full defection group can reach the
lower limit [bottom row in Fig.~\ref{fig2}]. However, the
multiplication factor within the groups along the boundary of
cooperators and defectors reaches a value between the lower and
upper limit.

In combination with the above investigations, let us now explain the
emergent results. Indeed, a feedback mechanism is at work between
the strategy distribution and the distribution of multiplication
factors in all the groups. Therefore, cooperators form compact
clusters, and these clusters can become larger and larger,
especially when the global cooperation level is not very high.
Meanwhile, the multiplication factor in the cooperators' clusters
becomes larger and larger, which provides cooperators, especially
the ones on the boundary, with a higher payoff. Whereas defectors
also gather together, and the multiplication factor in the
defectors' clusters becomes smaller and smaller [bottom row in
Fig.~\ref{fig2}]. In those interacting groups where the
multiplication factor is negative, it is better for the players to
choose the defective strategy such that they can have relatively
higher payoffs. In a sense, this adaptive mode can induce a
double-edged sword effect on the evolution of cooperation. However,
under the social learning defectors are inclined to learn from their
neighboring cooperators. As a consequence, the evolution of
cooperation can be favored by this locally adaptive investment
return.

If there is no limitation for the adaptive multiplication factor,
due to the continuing negative feedback effects cooperators cannot
invade the defectors' clusters, even if the feedback strength is
high. In this situation, cooperators can thrive, but cannot dominate
the whole population. When the limitation is considered for the
dynamical multiplication factor, the feedback mechanism, especially
the negative feedback effects, can be effectively weakened. If there
is too much restriction for the adaptive multiplication factor, that
is, $R$ is not very large, e.g., $R=4$, the multiplication factor in
some groups can reach the upper limit from a negative value due to
the social learning, but this upper limit cannot warrant the
promotion of cooperation in spatial PGG even if the feedback
strength is enough high. Whereas for a larger $R$, e.g., $R=10$, the
multiplication factor in the group along the boundary reaching the
upper limit can warrant a better promotion of cooperation. Hence,
this adaptive and bounded mode for the multiplication factor can
provide a better environment for the evolution of cooperation.

\begin{figure}[!ht]
\begin{center}
\includegraphics[width=5in]{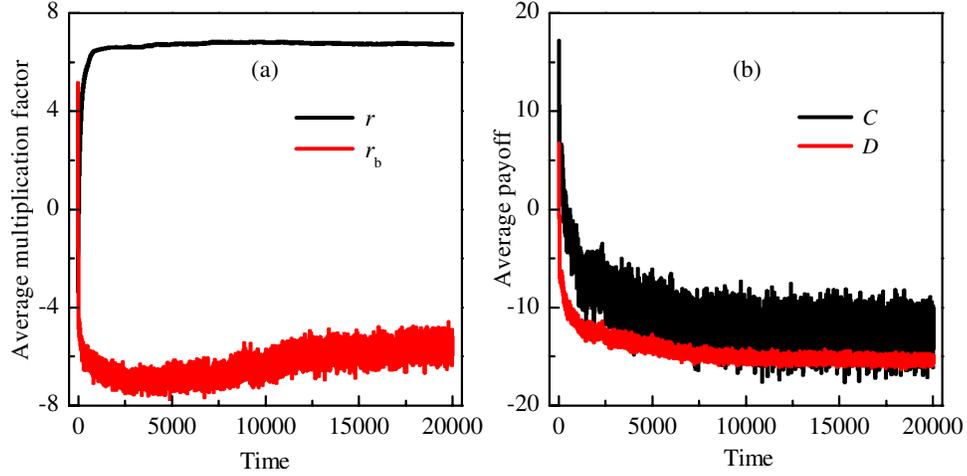}
\end{center}
\caption{ {\bf Time evolution of average multiplication factor and
payoffs.} Panel (a) depicts the time evolution of average values of
multiplication factor in the whole population and in the boundary
groups, respectively. Panel (b) depicts the time evolution of
average payoffs of cooperators and defectors along the boundary,
respectively. It can be observed that although the average value of
multiplication factor in the whole population is large enough for
the evolution of cooperation \cite{PRE09Szolnokia}, the average
value along the boundary becomes negative. Correspondingly, the
average payoffs of cooperators and defectors along the boundary are
both less than zero. As time increases, the average payoff of
cooperators along the boundary is a little higher than that of
defectors, but has larger fluctuations. Here, $\alpha=5$, $r_0=1$,
and $R=10$.} \label{fig3}
\end{figure}

However, under this adaptive and bounded mode, the multiplication
factor in some groups along the boundary cannot rapidly become a
positive and large value from a negative one when the feedback
strength is not very high [bottom row in Fig.~\ref{fig2}]. Although
the average multiplication factor $r$ in the whole population can
reach an enough high value which can make the cooperation level
reach one in the traditionally spatial PGG \cite{PRE09Szolnokia},
the average multiplication factor along the boundary of cooperators
and defectors $r_b$ becomes negative as time increases
[Fig.~\ref{fig3}(a)]. This does not provide a favorable environment
for players' interactions. Correspondingly, the average payoffs of
cooperators and defectors along the boundary are both negative.
Moreover, as time increases the average payoff of cooperators along
the boundary is just a litter higher than the one of defectors along
the boundary, but have larger fluctuations [Fig.~\ref{fig3}(b)].
Under the stochastic strategy updating, defectors do not always
successfully imitate their neighboring cooperators, but sometimes
may spread their strategy to the cooperators. As a result,
cooperators cannot defeat those defectors along the boundary, and
they can only coexist with defectors for an exceedingly long time.
On the contrary, when the feedback strength is high, by means of
social learning the multiplication factor in the group along the
boundary can have the opportunity to suddenly reach the upper limit,
which can warrant the invading of cooperative behavior into the
defectors' clusters. Finally, full cooperation can be achieved.

\begin{figure}[!ht]
\begin{center}
\includegraphics[width=5in]{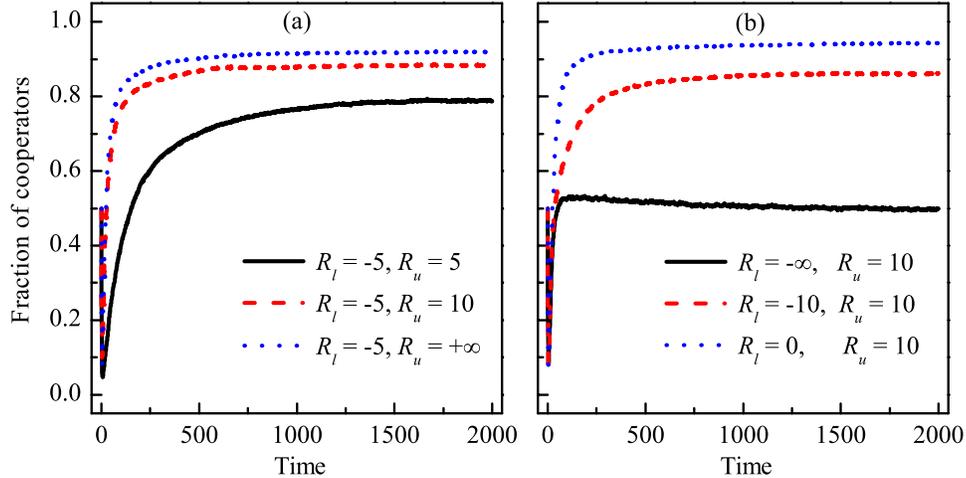}
\end{center}
\caption{ {\bf Cooperation promoted when the values of the lower and
upper limits of the investment returns are increased.} Panel (a)
depicts the fraction of cooperators in the whole population as a
function of time for fixed lower limit $R_l=-5$ and different values
of upper limit. Panel (b) depicts the fraction of cooperators in the
whole population as a function of time for fixed upper limit
$R_u=10$ and different values of lower limit. Increasing the values
of lower and upper limit can provide more positive effects on the
evolution of cooperation. Here, $\alpha=5$ and $r_0=1$.}
\label{fig4}
\end{figure}

In what follows, we study how cooperation evolves if the lower limit
is not the opposite number of the upper limit. Figure~\ref{fig4}(a)
shows the typical time evolution of cooperation for fixed $R_l=-5$
and three different values of $R_u$. We find that increasing the
value of $R_u$ can make the system reach a higher cooperation level,
but the cooperation level at equilibrium for $R_u=+\infty$ is just
slightly larger than the one for $R_u=10$. In Fig.~\ref{fig4}(b) we
show the fraction of cooperators as a function of time for fixed
$R_u=10$ and three different values of $R_l$. We find that
increasing the value of $R_l$ can make the system reach a higher
cooperation level. Moreover, as time increases the fraction of
cooperators first drops and then rapidly increases, but the larger
values of $R_u$ or $R_l$ make the cooperation level increases
faster. We also find that to have a favorable cooperation level, it
is better to set the lower limit higher and it is not necessary to
set the upper limit too high.

\begin{figure}[!ht]
\begin{center}
\includegraphics[width=5in]{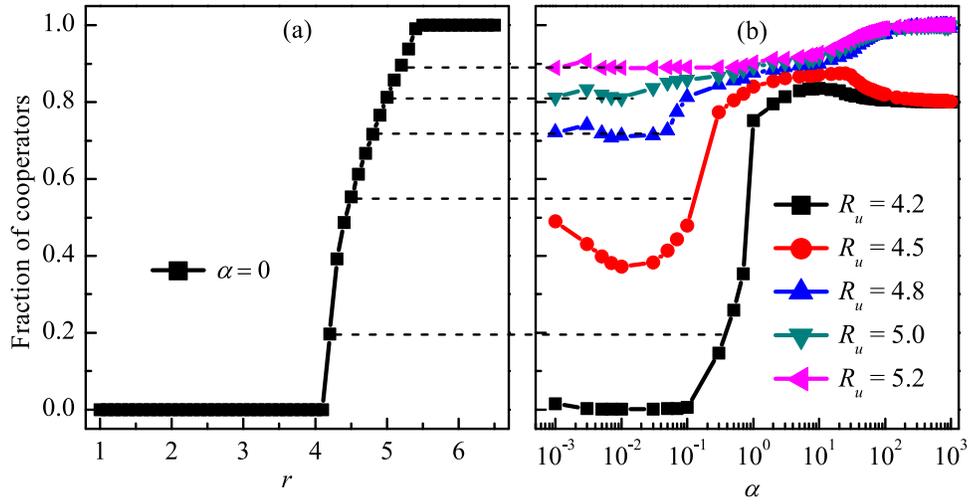}
\end{center}
\caption{ {\bf Cooperation promoted even when the investment return
is updated within the interval $[1, N]$.} Panel (a) shows the
fraction of cooperators as a function of $r_0$ for $\alpha=0$. In
this situation, the model recovers to the traditionally spatial PGG,
where the multiplication factor in each group is fixed at $r_0$ and
$r=r_0$. For $\kappa=1.0$, cooperators can survive only if
$r_0>4.1$, and they can dominate the whole population only if
$r_0>5.4$. Panel (b) shows the fraction of cooperators as a function
of $\alpha$ for fixed $R_l=1$ and different values of $R_u$.
Initially, the multiplication factor in each interacting group is
$r_0=R_u$. Dash lines are used to indicate the critical value of
$\alpha>0$ for a better promotion of cooperation in this adaptive
and bounded mode for the enhancement factor.} \label{fig5}
\end{figure}

Finally, we study whether cooperation can be better promoted if the
adaptive multiplication factor is constrained between $1$ and $N$ by
means of a comparative investigation. Previous work has reported
that in the traditional PGG where only one invariable multiplication
factor exists in the whole population, defectors outperform
cooperators in any given mixed group for $r<N$ \cite{JTB02Hauert}.
We further find that in the traditionally spatial PGG, for noise
value $\kappa=1.0$ cooperators can dominate the whole population
only if $r>5.4$, and they can survive in the system only if $r>4.1$,
as shown in Fig.~\ref{fig5}(a). It is worth pointing out that the
traditionally spatial PGG corresponds to the situation of $\alpha=0$
in this present model, where the multiplication factor in each
interacting group is fixed at $r_0=r$. In Fig.~\ref{fig5}(b), we set
$r_0=R_u$, and show the cooperation level as a function of feedback
strength $\alpha>0$ for $R_l=1$ and different values of $R_u>4.1$.
We see that the cooperator density varying with $\alpha$ displays
two different behaviors: for smaller values of $R_u$, e.g.,
$R_u=4.2$, the cooperation level first decreases and then increases
until reaching the maximum value. Subsequently, it decreases very
slowly with increasing $\alpha$ and its value approaches $0.8$; for
larger values of $R_u$, e.g., $R_u=4.8$, the cooperation level does
not change too much for small values of $\alpha$, then monotonously
increases to one with increasing $\alpha$. In addition, for smaller
values of $R_u$ just a small amount of $\alpha$ ($\alpha<0.5$) is
needed to warrant a better promotion of cooperation in comparison to
the traditionally spatial PGG, and the critical amount of $\alpha$
becomes smaller if the $R_u$ is increased. For larger values of
$R_u$, the cooperation level for any value of $\alpha>0$ is not less
than the one for $\alpha=0$. In fact, the average multiplication
factor in the population for $\alpha>0$ is not larger than the one
for $\alpha=0$, but these results suggest that cooperation can be
better promoted in comparison to the traditionally spatial model.
In addition, Fig.~\ref{fig5}(b) shows that the
cooperation level increases with increasing the upper limit of the
multiplication factor and the initial value of the multiplication
factor. We have also verified that increasing the initial fraction of
cooperators is beneficial for the evolution of cooperation in this
model.

\section*{Discussion}
In summary, we have presented a coevolutionary rule where the
multiplication factor in each interacting group is updated based on
the local strategy distribution in the group and the global strategy
distribution in the whole population, and studied its impact on the
evolution of cooperation in the spatial public goods game. We found
that this adaptive rule for the multiplication factor can
effectively enhance the evolution of cooperation. When the
appropriate bounded limitation for the dynamical multiplication
factor is further considered, cooperation can be better promoted. In
particular, full cooperation can be achieved in the system when the
feedback strength is high enough. Also, increasing the lower and
upper limit values of the multiplication factor is favorable for the
evolution of cooperation, but high cooperation level can be reached
even if the upper limit is not very large. We further found that
even if the multiplication factor is constrained to change between
one and the group size, cooperation can be better promoted in the
adaptive mode, in comparison to the classically spatial public goods
game where the payoff parameter in each group is fixed and
identical.

The adaptive mode for the investment returns results in that a
feedback mechanism is at work, that is, the Matthew effect is
introduced. From the viewpoint of this emergent feature, our model
is related to the one proposed by Perc \cite{PRE11Perc}, who
considered that the reproductive success of each individual is
updated by means of the enforcement of strategy and the distribution
of public goods is driven by the reproductive success of
individuals. Under this success-driven mechanism, cooperation can be
promoted. However, defectors can have a much higher payoff even in
the sea of cooperators, and easily enforce their strategy choice to
their neighbors. Correspondingly, the superpersistent defector
emergences spontaneously, and cooperators cannot dominate the whole
population. The complete dominance of cooperators is elusive even if
the limitation factor about the value of reproductive success is
considered. Whereas in our model, the multiplication factor in each
interacting group is updated based on the local and global strategy
distribution, which characterizes the local investment environment
for collective interactions, rather than individual's personality.
Also, we incorporate the limitation factor for the dynamical
multiplication factor. In this framework, cooperators and defectors
can form their own compact clusters respectively, and
correspondingly cooperators along the boundary can have a higher
payoff than the neighboring defectors. Under the social learning,
cooperators can easily spread their strategy even if the noise level
for strategy updating is large. In particular, when the feedback
strength is high, the interacting environment including some
defectors can rapidly become favorable. Thus, cooperators can
gradually invade defector's clusters, and finally dominate the whole
population. It could be concluded that our work further enriches the
knowledgeless of coevolutionary rules in PGGs, and importantly our
spatial PGG model with adaptive and bounded investment returns not
only can promote cooperation, but also make cooperators completely
dominate the population.

It is worth emphasizing the bounded values of the multiplication
factor play a different role in the evolution of cooperation in our
model in comparison to the one in Ref. \cite{PRL11Lee}. It is found
that the main results remain qualitatively if the limit value of the
payoff parameter in the prisoner's dilemma is large enough.
Moreover, it is demonstrated that the final cooperation level
strongly depends on the values of initial payoff parameter and
feedback strength in Ref. \cite{PRL11Lee}. In particular, with
increasing the initial payoff parameter, the probability that the
system ends in full cooperation state decreases. Whereas in this
work, we find that the introduced limitation factor can weaken the
Matthew effect, particularly the negative feedback effect on the
unfavorable interacting groups, which makes the limit value a
crucial model parameter. The limited negative effects can be
overcome via social learning. Hence, appropriate limitation can
warrant the best promotion of cooperation. In addition, we show that
the finial cooperation level not only depends on the the values of
initial payoff parameter and feedback strength, but also depends on
the limit values. With increasing the initial payoff
parameter and the initial fraction of cooperators, cooperation can
be better enhanced when the dynamical multiplication factor is
limited. In a sense, this work further explores the effects of
adaptive and bounded game payoffs on the evolution of cooperation.

During the coevolutionary process, the investment return in most of
interacting groups reaches the upper or lower limit due to the
feedback effects. This segregation and polarization of investment
returns occurs spontaneously over time, which is different from the
distribution in Ref. \cite{EPL10Shi}. In the latter case, the
distribution of the multiplication factor is artificially introduced
by the authors and does not change during the evolutionary process.
Although the emergent values of the multiplication factor in all the
interacting groups do not display too much diversity, we find that
cooperation can be promoted in this adaptive and bounded mode.
Compared with the results in the traditionally spatial PGG model,
cooperation can be better promoted even if the multiplication factor
can only change between one and the group size.

In the present model, we consider the adaptive mode for the
multiplication factor in a group based on the local cooperation
level in the classical PGG where players just have two discrete
strategy choices $C$ or $D$, and correspondingly the local
cooperation level only has several finite values. To make the local
cooperation level change continuously between zero and one, we also
introduce the adaptive and bounded investment returns into the
spatial continuous PGG \cite{JTB06Janssen,JTB08Kamimura,
JTB11Cressman}, and still find that this PGG with adaptive and
bounded investment returns promotes cooperation. We
also test our model in well-mixed populations as well as on other types
of interactions networks, and still find that cooperation can be enhanced by the proposed coevolutionary rule. Moreover, we would like to point out that in this work we fix the value of noise to one. In general, the qualitative
behavior of the system remains unchanged for other values of noise,
although for pairwise interactions there may exist an optimal value of noise at which the evolution of cooperation is most successful \cite{PRE09Szolnokia}.
It could be inferred that if we can choose the optimal noise value
for strategy updating, the positive effects from social learning can
be amplified. We also believe that, cooperation can be better
promoted if we further incorporate the selection of noise level in
strategy adoption \cite{EPL09Szabo, PRE09Szolnokib} into this
adaptive and bounded mode for the multiplication factor.

\section*{Methods}

We consider the PGG on a square lattice of size $L\times L$ with
periodic boundary conditions. Each individual who is a pure
strategist can only follow two simple strategies: cooperate ($C$)
and defect ($D$). Cooperators contribute a fixed amount (here
considered to be equal to $1$ without loss of generality) to the
public good while defectors contribute nothing. The sum of all
contributions in each group $i$ is multiplied by the factor $r_i$,
and the resulting public goods are distributed among all the group
members. Correspondingly, the payoff of player $x$ from the group
$i$ is

\begin{equation}
P_x^i=\left\{
\begin{array}{lll}
r_i\frac{n_i}{N}-1 & \mbox{ if }s_x=C,\\
r_i\frac{n_i}{N} & \mbox{ if }s_x=D,
\end{array} \right.
\end{equation}
where $s_x$ denotes the strategy of player $x$, $n_i$ denotes the
number of cooperators in the group $i$, and $N$ denotes the group
size. Here, we consider the square lattice with von Neumann
Neighborhood. Accordingly, the interacting group size is fixed at
$N=5$, and each individual belongs to five different groups. The
payoff of each player is accumulated from the fixed five interacting
groups, and thus player $x$' total payoff $P_x=\sum{_{i}} P_x^i$.

After playing the games, the multiplication factor in each group
needs to be updated. Specifically, we assume that the multiplication
factor of the group centered on player $x$ at time $t+1$ is
\begin{equation}
r_x(t+1)=r_x(t)+\alpha[\rho_x(t)-\rho(t)],
\end{equation}
where $\alpha$ controls the strength of feedback from the comparison
between the local and global cooperative environments, $r_x(t)$ is
the multiplication factor of the focal individual $x$'s group at
time $t$, $\rho(t)$ is the fraction of cooperators in the whole
population at time $t$, and $\rho_x(t)$ is the local cooperation
level in the group centered on $x$ at time $t$. Here,
$\rho_x(t)=n_x/N$, where $n_x$ denotes the number of cooperators in
the group where player $x$ is the focal individual at time $t$.
Moreover, we consider the limitation for the adaptive multiplication
factor following previous work \cite{PRL11Lee}, that is, letting
$r_x(t+1)=R_u$ if $r_x(t+1)>R_u$ and letting $r_x(t+1)=R_l$ if
$r_x(t+1)<R_l$. Here, $R_u$ and $R_l$ respectively represent the
upper and lower limits of the multiplication factor in each group,
and they satisfy the following inequalities: $R_u>R_l$ and $R_u>1$.
In particular, when $R_u=-R_l=R$, the multiplication factor is
constrained in a symmetric interval, which is the same to the
setting in Ref.\cite{PRL11Lee}. Moreover, when $R=+\infty$ (no
bounded limitation for the multiplication factor) or $\alpha=0$ (no
updating for the multiplication factor), the average multiplication
factor in all the interacting groups $r=L^{-2}\sum _x r_x(t)=r_0$,
where $r_0$ is the initial value of the multiplication factor in
each group.

Subsequently, each player is allowed to learn from one of its
neighbors and update its strategy. Player $x$ adopts the randomly
chosen neighbor $y$' strategy with a probability depending on the
payoff difference as
\begin{equation}
f(P_y-P_x)=\frac{1}{1+\exp[-(P_y-P_x)/\kappa]},
\end{equation}
where $\kappa$ denotes the amplitude of noise \cite{PRE98Szabo},
accounting for imperfect information and errors in decision making.
Following previous work \cite{PRL11Lee}, we simply set $\kappa=1.0$
representing that it is very likely that the better performing
players will pass their strategy to other players, yet it is
possible that players will occasionally learn also from the less
successful neighbors.

Simulations of this spatial PGG model are performed by means of a
synchronous updating rule, using $L=100$ to $400$ system size.
Initially, the two strategies of $C$ and $D$ are randomly
distributed among the population with an equal probability, and the
multiplication factor in each interacting group has the same value
$r_0$. The key quantity for characterizing the cooperative behavior
of the system is the density of cooperators, which is defined as the
fraction of cooperators in the whole population. The system can
reach a dynamical equilibrium after a suitable transient time
\cite{JTB08Gomez-Gardenes,PRE08Vukov,PRE08Assenza}. Then the density
of cooperators reaches its asymptotic value $\rho$ and remains there
within small fluctuations (less than $0.01$). This asymptotic value
is taken to describe the cooperation level in the whole population,
and all the simulation results are averaged over $100$ different
realizations of initial conditions.

\section*{Acknowledgments}
YL acknowledges the financial support from the Special Fund for the Basic Scientific Research of Central Colleges, Chang'an University, the Special Fund for the Basic Research Support Planning of Chang'an University, and the open fund of road traffic intelligent detection and equipment engineering research center, Shaanxi province (Grant
No. CHD2010JC134). YZ acknowledges the financial support from the National Natural Science Foundation of China (NSFC) (Grand No. 11161011). LW acknowledges the financial support from 973 Program (2012CB821203) and the National Natural Science Foundation of China (NSFC) (Grant Nos. 61020106005 and 10972002). MP acknowledges the financial support from the Slovenian Research Agency (ARRS) (Grant J1-4055). The funders had no role in study design, data collection and analysis, decision to publish, or preparation of the manuscript.

\clearpage

\end{document}